\documentclass{article}
\usepackage{graphicx} 
\def\lesssim{\ {\raise-.5ex\hbox{$\buildrel<\over\sim$}}\ }
\def\kms{kms$^{-1}$\,}

\begin{document}

\begin{center}
\noindent\huge{Tidally Trapped Pulsations in a close binary star system discovered by TESS}\\
\end{center}

\noindent\large{G. Handler,$^{1}$ D. W. Kurtz,$^2$ S. A. Rappaport,$^3$ H. Saio,$^4$ J. Fuller,$^{5}$ D. Jones,$^{6,\,7}$ Z. Guo,$^8$ S. Chowdhury,$^1$ P. Sowicka,$^1$ F. Kahraman Ali\c{c}avu\c{s},$^{1,\,9}$ M. Streamer,$^{10}$ S. J. Murphy,$^{11}$ R. Gagliano,$^{12}$ T. L. Jacobs$^{13}$ and A. Vanderburg\,$^{14,\,15}$}\\


\noindent Nicolaus Copernicus Astronomical Center, Polish Academy of Sciences, ul. Bartycka 18, 00-716, Warsaw, Poland\\
$^{2}$ Jeremiah Horrocks Institute, University of Central Lancashire, Preston PR1 2HE, UK\\
$^3$ Department of Physics, and Kavli Institute for Astrophysics and Space Research, M.I.T., Cambridge, MA 02139, USA\\
$^4$ Astronomical Institute, Graduate School of Science, Tohoku University, Sendai 980-8578, Japan\\
$^{5}$ Division of Physics, Mathematics and Astronomy, California Institute of Technology, Pasadena, CA 91125, USA\\
$^6$ Instituto de Astrof\'isica de Canarias, E-38205 La Laguna, Tenerife, Spain\\
$^7$ Departamento de Astrof\'isica, Universidad de La Laguna, E-38206 La Laguna, Tenerife, Spain\\
$^8$ Department of Astronomy and Astrophysics, Pennsylvania State University, 421 Davey Lab, University Park, PA 16802, USA\\
$^{9}$ \c{C}anakkale Onsekiz Mart University, Faculty of Sciences and Arts, Physics Department, 17100, \c{C}anakkale, Turkey\\
$^{10}$ Research School of Astronomy and Astrophysics, Australian National University, Canberra, ACT, Australia\\
$^{11}$ Sydney Institute for Astronomy (SIfA), School of Physics, University of Sydney, NSW 2006, Australia\\
$^{12}$ Planet Hunters\\
$^{13}$ Amateur Astronomer, 12812 SE 69th Place Bellevue, WA 98006, USA\\
$^{14}$ Department of Astronomy, The University of Texas at Austin, 2515 Speedway, Stop C1400, Austin, TX 78712, USA\\
$^{15}$ NASA Sagan Fellow

\newpage

{\bf It has long been suspected that tidal forces in close binary stars could modify the orientation of the pulsation axis of the constituent stars. Such stars have been searched for, but until now never detected. Here we report the discovery of tidally trapped pulsations in the ellipsoidal variable HD~74423 in {\it TESS} space photometry data. The system contains a $\delta$~Scuti pulsator in a 1.6-d orbit, whose pulsation mode amplitude is strongly modulated at the orbital frequency, which can be explained if the pulsations have a much larger amplitude in one hemisphere of the star. We interpret this as an obliquely pulsating distorted dipole oscillation with a pulsation axis aligned with the tidal axis. This is the first time that oblique pulsation along a tidal axis has been recognized. It is unclear whether the pulsations are trapped in the hemisphere directed towards the companion or in the side facing away from it, but future spectral measurements can provide the solution. In the meantime, the single-sided pulsator HD~74423 stands out as the prototype of a new class of obliquely pulsating stars in which the interactions of stellar pulsations and tidal distortion can be studied.}

Much of the present-day understanding of the universe is rooted in the knowledge of the basic parameters and structure of the stars. The precise determination of these parameters rests on two methods: the analysis of eclipsing binary stars and asteroseismology. Whereas detached eclipsing binaries facilitate the determination of global stellar parameters to the highest precision (e.g., see ref.$^1$), asteroseismology allows the determination of interior stellar structure in fine detail (ref.$^2$). Thus, even more rewarding are asteroseismic analyses of the components of detached eclipsing binaries.

Proximity effects between binary star components can influence stellar oscillations. The modification of a star's interior structure due to mass transfer can be traced asteroseismically. Many decades ago it was theoretically predicted that the excitation of stellar oscillations can be augmented by time-varying tides (ref.$^3$), but convincing observational evidence has only been accumulated in the  recent past (ref.$^4$).

It was only with the advent of the {\it Kepler} space telescope that binary components whose oscillations were affected by time-varying tides were discovered in larger numbers. Most prominent among these are the highly eccentric binary ``Heartbeat'' stars (e.g., refs.$^{5,6,7,8}$). Some of these stars show tidally excited gravity modes that are exactly resonant with high harmonics of the orbital frequency. 

Aside from tidal excitation, it has been speculated (ref.$^9$) that a binary companion could cause a tilt of the stellar pulsation axis, which would result in amplitude modulation of a nonradial pulsation over the orbit. This is analogous to what had already been observed in the rapidly oscillating Ap (roAp) stars (ref.$^{10}$), where the pulsation axis is aligned with the stars' magnetic axis.

The roAp stars show high radial overtone, nonradial pulsation modes with frequency multiplets that have frequency separations exactly equal to the known rotation frequency for a particular Ap star, which is determined with precision from rotationally induced light variations caused by long-lived abundance spots. In the asymptotic regime, ref.$^{11}$ showed that when $n \gg \ell$, where $n$ is the radial overtone and $\ell$ is the spherical degree of a pressure (p)~mode, 

\begin{equation}
{\omega_{\ell,m} = \omega_{\ell,0} + m(1-C_{n,\ell})\Omega ,}
\end{equation}

\noindent where $\omega$ is the pulsation frequency, $\Omega$ is the rotation frequency and $C_{n,\ell}$ is the `Ledoux constant', which for p~modes usually differs from zero by only a few per cent. Ref.$^{10}$ was able to show for the roAp stars that $C_{n,\ell}$ is zero, if the above equation is applied, and proposed instead the oblique pulsator model, where the pulsation axis is the magnetic axis, which is known to be oblique to the rotation axis in most Ap stars. The roAp stars thus became the first pulsating stars where it could be shown that the pulsation axis was {\it not} the rotation axis. There were many subsequent developments of the oblique pulsator model, and it is now believed that the pulsation axis is neither the magnetic axis, nor the rotation axis, but rather lies along a plane between those (ref.$^{12}$).

With it being clear from the roAp stars that non-radial pulsators can have pulsation axes other than the rotation axis, an obvious idea is that in close binary stars, the tidal distortion may not only tilt the pulsation axis as suggested by ref.$^{1}$, but even be sufficient to make the line of apsides the pulsation axis. Refs.$^{13, 14}$ also considered this possibility. Searches for this new type of obliquely pulsating stars have been ongoing with high precision {\it Kepler} mission data, and now with the new {\it TESS} data. However, no pattern of frequencies typical of oblique pulsation, as in the roAp stars, was found - until now.

HD~74423 is a $V=8.61$ mag A-type star in the Southern Hemisphere for which there is little information in the literature.  Ref.$^{15}$ gave a spectral type of A1V. Four decades later, ref.$^{16}$ reported photometric variability with a period of 0.79037(1)\,d and a peak-to-peak amplitude of 0.08 mag found in ASAS-3 data (ref.$^{17}$). On this basis, HD~74423 was classified as a candidate photometrically variable chemically peculiar star. 

Ref.$^{18}$ reported HD~74423 as a chemically peculiar star of the $\lambda$~Bootis type, with a spectral type of A7V\,kA0mA0, and determined $T_{\rm eff}=8100$\,K, log\,$g=3.6$, $[M/H]=-1$ and $E(B-V)=0.055$. The star's Gaia distance is $491 \pm 8$\,pc (ref.$^{19}$), hence $L=83 \pm 4\,L_\odot$. High-resolution spectroscopy of HD 74423 (Supplementary Figure 1) shows that the most prominent spectral lines are double, with a velocity separation of about 150\,\kms, and that both components share the $\lambda$~Bootis-type spectral peculiarity. Assuming that they have roughly the same luminosity, component masses of $M \approx 2.25\,M_\odot$ and an age of $t \approx 8\times10^8$ yr for the system are estimated from evolutionary tracks of the rotating models in ref.$^{20}$.

The peculiar pulsation pattern in HD 74423 = TIC 355151781 was first noticed as being unusual by three of us (R.\,G., T.\,J and S.\,J.\,M.) during visual surveys of the {\em TESS} light curves from Sector 9. Such visual surveys have led to the discovery of a number of unusual and interesting astronomical objects not picked up by standard Box Least Squares (`BLS'; ref.$^{21}$), searches for periodic signals. Examples of these discoveries are listed in Sect.~2.2 of ref.$^{22}$. HD~74423 was ultimately observed by {\em TESS} during Sectors 9, 10, and 11 in 2-min cadence. 

\begin{figure}
\centering
\includegraphics[width=0.999\linewidth,angle=0]{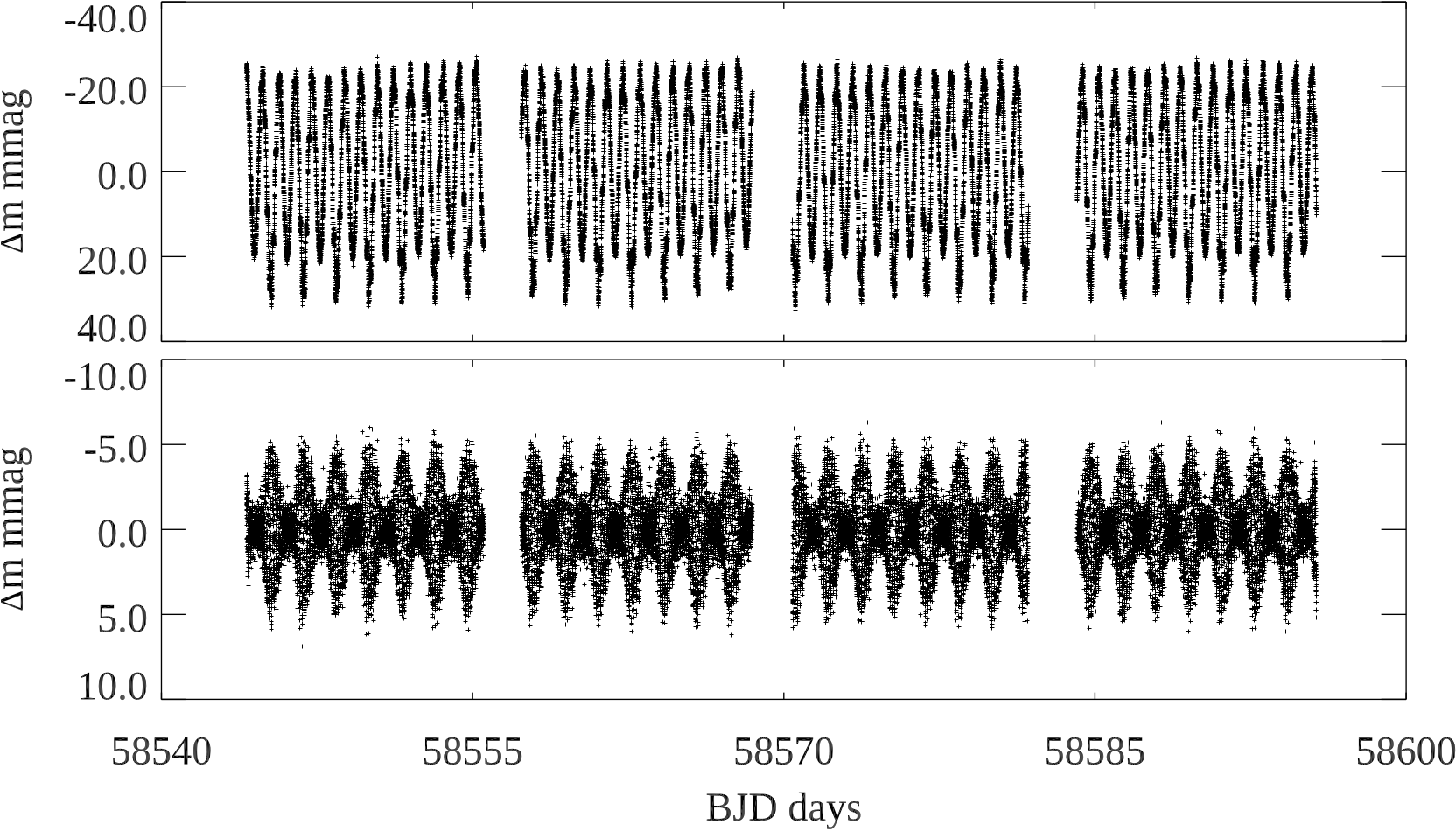}
\caption{{\it TESS} light curves of HD~74423. Top: A section of the light curve showing the clear ellipsoidal light variations, along with higher frequency pulsations. The sections of the light curve not shown are similar. Bottom: The same section of the light curve after pre-whitening the orbital variations and low frequency artefacts. The modulation of the pulsation amplitude with the orbital period is striking.}
\label{fig:lc1}
\end{figure}

The top panel of Fig.~1 shows a section of the initial light curve (the full light curve would be too compressed at this scale to see the details) where the variations already reported by ref.$^{16}$ are obvious. However, the light curve shows minima of alternating depth whereas the maxima do not alternate. This type of light curve is characteristic of an ellipsoidal variable (ref.$^{23}$) and less reminiscent of a rotational variable as implied by ref.$^{16}$. The differing minima arise, in part, from differential gravity darkening near the L1 and L2 points of the tidally distorted star. In the lower panel of Fig.\,1, the orbital variability has been removed, which shows that the amplitude of the pulsation is clearly modulated with the orbit. 

The amplitude changes by about a factor of 10 between the two ellipsoidal light minima. Obscuration of the pulsating star by the companion cannot explain this modulation, as it would also cause eclipses of at least 0.7 mag depth, which are not seen. We therefore face the curious situation that the pulsations have much larger amplitude on one hemisphere of the star.

The top panel of Fig.\,2 shows the initial amplitude spectrum, where the second harmonic of the orbital frequency is the highest peak, as expected for a double-wave, ellipsoidal light curve in a close binary. The derived orbital frequency is $\nu_{\rm orb} = 0.6326218 \pm 0.0000006$\,d$^{-1}$ ($P_{\rm orb} = 1.580723 \pm 0.000002$\,d). 

The second panel  of Fig.\,2 shows the amplitude spectrum of the residuals after a five-harmonic fit of the orbital variation $\nu_{\rm orb}$ has been removed from the data. The pulsation multiplet and two of its harmonic multiplets are visible in this panel, as are low frequency peaks that are instrumental artefacts. We removed those low frequency artefacts with a high-pass filter to produce the amplitude spectrum in the third panel, which shows only the pulsation frequency and harmonics along with their orbital sidelobes.

\begin{figure}
\centering
\includegraphics[width=0.88\linewidth,angle=0]{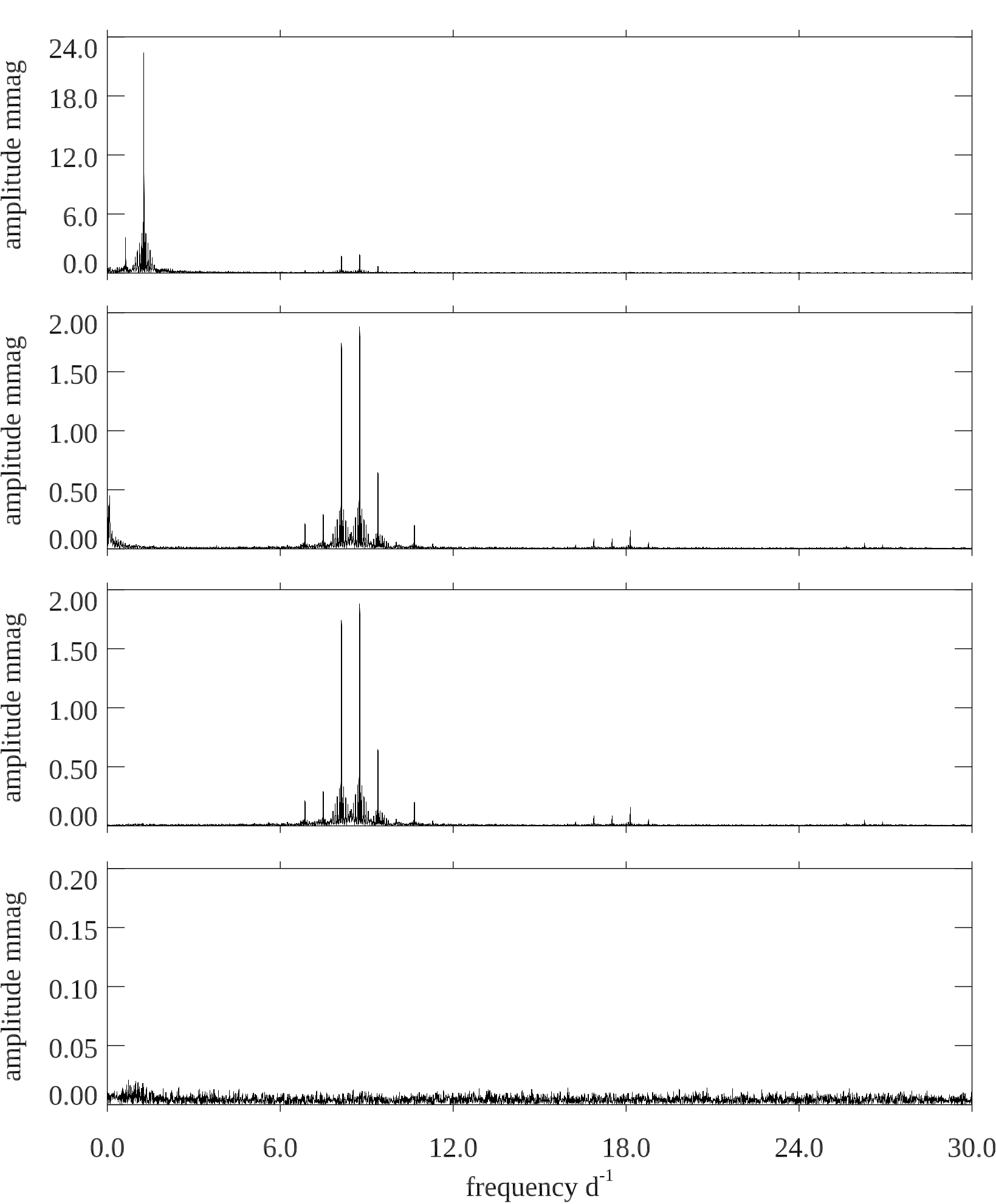}	
\caption{Fourier amplitude spectra of the {\it TESS} photometry of HD 74423. Top: The initial Fourier amplitude spectrum of the light curve, where the highest peak is at twice the orbital frequency, $\nu_{\rm orb} = 0.6326218 \pm 0.0000006$\,d$^{-1}$. Second panel: The amplitude spectrum of the residuals after pre-whitening a harmonic series of five terms based on the orbital frequency. It can be seen that there are low frequency peaks, which are instrumental artefacts. The pulsation multiplet centred around $\nu_1 = 8.756917 \pm 0.000010$\,d$^{-1}$ and its harmonics are visible. The third panel shows the same as the second, but after a high-pass filter has removed the low-frequency artefacts. The multiplets can easily be seen. The bottom panel shows the amplitude spectrum of the residuals after the fit shown in Table\,2. Note the changes in ordinate scale.}
\label{fig:ft1}
\end{figure}

From the third panel in Fig.\,2, it is apparent that most of the pulsational variation is represented by a frequency multiplet centred on the highest peak. By extracting the frequencies in this multiplet and examining their frequency separations, it is apparent that all are separated within the errors by the orbital frequency, $\nu_{\rm orb} = 0.6326218 \pm 0.0000006$\,d$^{-1}$. There is, therefore, only one pulsation frequency, $\nu_1 = 8.756917 \pm 0.000010$\,d$^{-1}$, and its orbital sidelobes and harmonics. The amplitude of signal $\nu_1 - \nu_{\rm orb}$ is similar to that of $\nu_1$ itself which makes it another candidate to be the actual pulsation frequency. However, the presence of the third harmonic triplet that is symmetric around $3\nu_1$ clearly argues against that hypothesis.

The harmonic multiplets are likewise separated by $\nu_{\rm orb}$. We therefore used a combination of linear and nonlinear least-squares fitting to optimise our determinations of the frequencies, amplitudes and phases for the multiplets. Table\,2 shows the results of those procedures. The phases of the pulsation frequency and its first sidelobes are very close to equal at the time when the line of sight is along the orbital line of apsides. This is a clear signature of oblique pulsation along that axis.  We note that the frequency ratio of the pulsation frequency to the orbital frequency is $13.84226 \pm 0.00003$, which is nearly 6000\,$\sigma$ away from an integer, hence the pulsation is unlikely tidally excited.

\begin{figure}
\centering
\includegraphics[width=0.999\linewidth,angle=0]{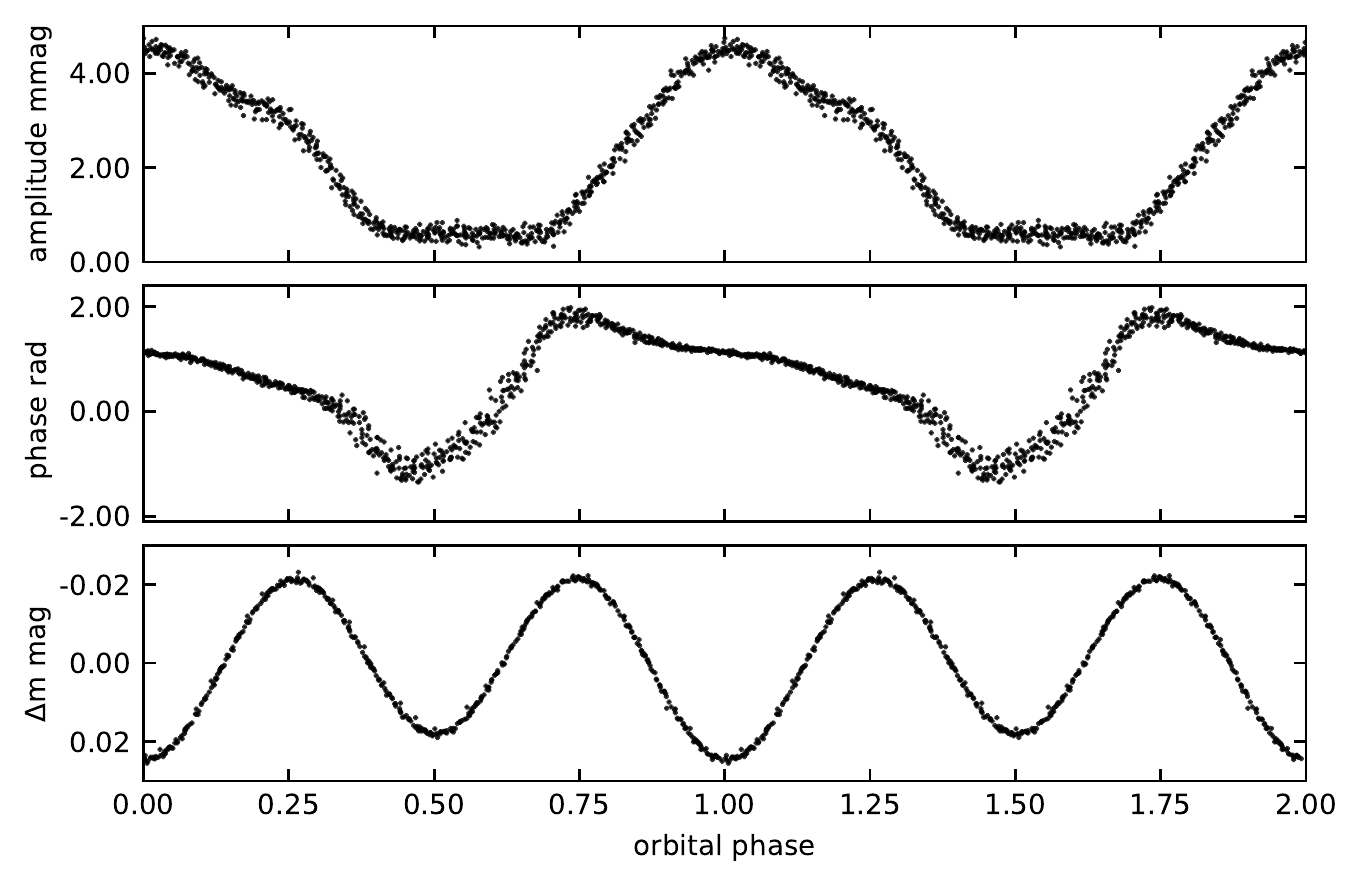}	
\caption{Run of the pulsation amplitude and phase over the orbital period. Top: The pulsation amplitude variation as a function of orbital phase taking the $8.756917$\,d$^{-1}$ frequency to be the pulsation frequency. The zero point in time is $t_0 = {\rm BJD}\,2458584.78684$, which was chosen to set the two first orbital sidelobes to have equal phase. That coincides with pulsation amplitude maximum, as expected for an oblique pulsator.  Middle: the pulsation phase variation over the orbital cycle with a range of $2\pi$ radians.  The quicker phase reversals come at amplitude minimum. Bottom: the orbital light variations as a function of orbital phase for comparison. The data have been binned to 200-s, and the pulsation variations have been removed. It can be seen that orbital light minimum coincides with pulsation maximum, as expected in the oblique pulsator model.}
\label{fig:phamp}
\end{figure}

\setcounter{table}{0}
\begin{table}
\centering
\caption{A least squares fit of the frequency multiplets for $\nu_1$ and its harmonics. The zero point for the phases, $t_0 = {\rm BJD}~2458584.78684$,  has been chosen to be a time when the two first orbital sidelobes have equal phase; note that those are both $-1.833$\,rad, and the phase of $\nu_1$ is close to these. It can be seen that (Fig. 3) orbital light minimum coincides with pulsation maximum, as expected in the oblique pulsator model.}
\begin{tabular}{rrrr}
\hline
&\multicolumn{1}{c}{frequency} & \multicolumn{1}{c}{amplitude} &   
\multicolumn{1}{c}{phase}  \\
&\multicolumn{1}{c}{d$^{-1}$} & \multicolumn{1}{c}{mmag} &   
\multicolumn{1}{c}{radians}   \\
& & \multicolumn{1}{c}{$\pm 0.004$} \\
\hline
 $\nu_1 - 5\nu_{\rm orb}$ & $5.593807 $ & $0.032 $ & $-1.100 \pm 0.119$ \\
 $\nu_1 - 4\nu_{\rm orb}$ & $6.226429 $ & $0.017 $ & $1.736 \pm 0.216$ \\
 $\nu_1 - 3\nu_{\rm orb}$  & $6.859051 $ & $0.228 $ & $1.985 \pm 0.017$ \\
 $\nu_1 - 2\nu_{\rm orb}$ & $7.491673 $ & $0.292 $ & $-1.751 \pm 0.013$ \\
 $\nu_1 - \nu_{\rm orb}$ & $8.124295 $ & $1.757 $ & $-1.833 \pm 0.002$ \\
 $\nu_1$ & $8.756917 $ & $1.894 $ & $-2.129 \pm 0.002$ \\
 $\nu_1 + \nu_{\rm orb}$ & $9.389539 $ & $0.656 $ & $-1.833 \pm 0.006$ \\
 $\nu_1 + 2\nu_{\rm orb}$ & $10.022161 $ & $0.057 $ & $1.427 \pm 0.066$ \\
 $\nu_1 + 3\nu_{\rm orb}$ & $10.654783 $ & $0.195 $ & $-1.524 \pm 0.019$ \\
 $\nu_1 + 4\nu_{\rm orb}$ & $11.287405 $ & $0.039 $ & $-1.562 \pm 0.096$ \\
 $\nu_1 + 5\nu_{\rm orb}$ & $11.920027 $ & $0.012 $ & $2.125 \pm 0.309$ \\
\hline
\hline
 $2\nu_1 - 2\nu_{\rm orb}$ & $16.248589 $ & $0.036 $ & $-0.077 \pm 0.105$ \\
 $2\nu_1 - \nu_{\rm orb}$ & $16.881211 $ & $0.090 $ & $-1.082 \pm 0.042$ \\
 $2\nu_1$ & $17.513833 $ & $0.086 $ & $2.381 \pm 0.043$ \\
 $2\nu_1 + \nu_{\rm orb}$ & $18.146455 $ & $0.157 $ & $2.467 \pm 0.024$ \\
 $2\nu_1 + 2\nu_{\rm orb}$ & $18.779077 $ & $0.060 $ & $2.555 \pm 0.062$ \\
 $3\nu_1 - \nu_{\rm orb}$ & $25.638128 $ & $0.025 $ & $-1.822 \pm 0.151$ \\
 $3\nu_1$ & $26.270750 $ & $0.053 $ & $-1.482 \pm 0.071$ \\
 $3\nu_1 + \nu_{\rm orb}$ & $26.903372 $ & $0.038 $ & $-1.638 \pm 0.098$ \\
\hline
\hline
\end{tabular}
\label{table:2}
\end{table}

Finally, the pulsation frequency $\nu_1 = 8.756917$\,d$^{-1}$ was fitted by least-squares to sections of the data 0.1-d long to examine the amplitude and phase variations over the orbital cycle. The original light curve was binned into 200-min time bins and all three curves are plotted in Fig.\,3. This figure shows that orbital light minimum and pulsation maximum coincide at the time when the line of sight is along the line of apsides.

There is, in principle, a small contribution to the multiplet sidelobes because of the frequency variation caused by the orbital motion (Doppler shift) known as frequency modulation (FM). The relative amplitudes of the first sidelobes compared to that of the central frequency for this effect is given by equation 21 of ref.$^{24}$ (note, however, that there is a typographical error in that equation; the first term should be $(2 \pi G)^\frac{1}{3}$/c). Applying the correct form of that equation gives their factor $\alpha \sim 0.001$. That is, the sidelobes generated by the frequency modulation are only 1/1000 the amplitude of the central peak, and this is negligible in our case here -- being only half of the error in the amplitude determination.

We have modelled the variation of the pulsation amplitude and phase over the orbit in analogy to magnetically modulated variability in the roAp stars. We assume luminosity variations arise from the temperature variation $\delta T$ on the stellar surface, which can be decomposed into spherical harmonics
\begin{equation}
\delta T \propto e^{i\omega t}\sum_{\ell=0}^2 A_\ell Y_\ell^0(\theta_{\mathrm p},\phi_{\mathrm p}),
\end{equation}
where $(\theta{_\mathrm p},\phi_{\mathrm p})$ represent spherical coordinates whose axis aligns with the line of apsides; i.e., we assume the pulsation is axisymmetric about the tidal axis. Converting the coordinate $(\theta_{\mathrm p},\phi_{\mathrm p})$ to the coordinate whose axis is the rotation axis, and to the inertial coordinate whose axis is along the line of sight, and then integrating the visual hemisphere at each epoch, we obtain the luminosity variation as a function of time as
\begin{equation}
\Delta L(t) \propto  e^{i\omega t}\sum_{\ell=0}^2N_\ell A_\ell\sum_{m=-\ell}^{\ell}d^\ell_{m,0}(\beta)d^\ell_{0,m}(i_o)e^{-im\Omega t},
\end{equation}
where $N_\ell$ is defined in Eq. 23 of ref.$^{25}$, and the coefficients $d^\ell_{m,0}(\beta)$ and $d^\ell_{0,m}(i_o)$ arise in converting the coordinate $(\theta_{\mathrm p},\phi_{\mathrm p})$ to the one associated with the rotation axis (the pulsation axis is inclined to the rotation axis by $\beta$) and to the coordinate axis associated with the line of sight (the line of sight is inclined to the rotation axis by angle $i_o$; see e.g., refs.$^{25,26}$ for details).

We have assumed $\beta = 90^\circ$ for a pulsation axis in the orbital plane, and have also assumed an inclination angle $i_o=40^\circ$ due to the absence of eclipses in the light curve. We then searched for a set of amplitude ratios $(A_1,A_2)/A_0$ (which are complex numbers in general) to best reproduce the observed amplitude and phase modulation of HD~74423 (Fig.~3). The best-fit model was obtained with $A_1/A_0 = (-1.3, -0.4i)$ and $A_2/A_0 = (1.0, 0.0i)$. The amplitude and phase modulations as well as the amplitudes of sidelobes are compared with the observed ones in Fig.~4. Except at the orbital phases of 0.3 and 0.7, the phase and amplitude variability are reasonably reproduced, showing the pulsation is largely confined to one hemisphere.

\begin{figure}
\centering
  \includegraphics[trim=1.5cm 1.1cm 1cm 1.2cm,width=0.75\textwidth]{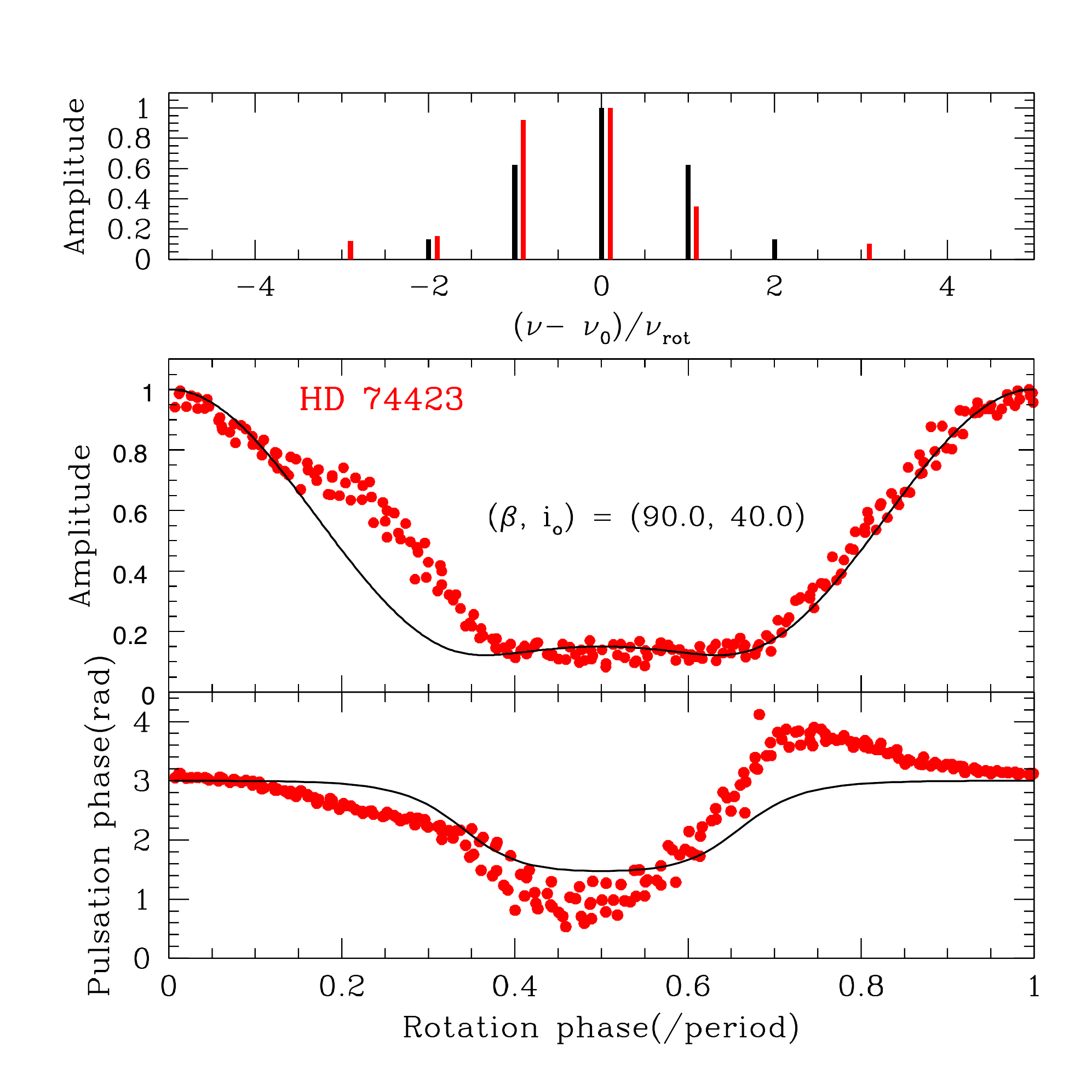}
    \caption{Top: comparison of the multiplet amplitudes (red lines) and those of our best-fit model (black lines) normalised by the amplitude of the central frequency. Middle: amplitude modulation of HD 74423 (red dots) compared with the model (black lines). Bottom: same comparison, but for the phase modulation.}
    \label{fig:orbfit}
\end{figure}

The deviations between our fit and the data cannot be avoided if we adopt an axisymmetric eigenfunction with respect to the pulsation axis, so it is clear that the pulsation is not exactly aligned with the tidal axis. The modelled surface amplitude distribution of the pulsation is shown in Supplementary Figure 2, again showing the higher amplitude near the tidal axis.

The chemical peculiarity of $\lambda$~Bootis stars is believed to stem from accretion of metal-depleted gas from either the interstellar medium or from a circumstellar disk or shell (ref.$^{27}$). Because searches for magnetic fields in $\lambda$~Bootis stars (ref.$^{28}$) have not yielded positive results to date, the current understanding is that, unlike the roAp stars, $\lambda$~Bootis stars do not have surface spots. Relative to the ephemeris $T_0= {\rm HJD}~ (2458584.7930 \pm 0.0004) + (1.580723 \pm 0.000002)\times E$, where $T_0$ is the time of the deeper light minimum and $E$ is the epoch, the orbital phase of the spectral observation is $\phi=0.7689\pm0.0004$, at $E=125$. This is close to the phase of maximum radial velocity separation of the two components if they were gravitationally bound, and indeed a 150\,\kms radial velocity difference is observed. According to Kepler's Third Law this is consistent with orbital motion of the components of HD~74423 seen at an aspect of $i\approx30^o$, and hence represents very strong evidence that the 1.58-d light variation is indeed orbital and not due to surface spots. In addition, there is no plausible explanation for why a spotted star would have its pulsations restricted to certain regions of the star.

The tidal synchronization time scale of the HD~74423 system was estimated following ref.$^{29}$ (Eqs. 4 and 5 for early-type stars with radiative envelopes), for a series of combinations of 2.0 M$_{\odot} < M_1=M_2 < 2.5$ M$_{\odot}$, with $T_{\rm eff1}=T_{\rm eff2}=8000$\,K. The resultant circularization time scales range from $4.0\times10^8 - 8.0\times10^8$\,yr, and the corresponding synchronization timescales are one order of magnitude smaller ($1.0\times10^7 - 2.6\times10^7$\,yr). The two stars are close to the end of the main sequence, and in all cases the synchronziation timescales are more than one order of magnitude shorter than the main sequence lifetime. Given the previous age estimate of HD~74423 ($8.0\times10^8$\,yr) it is thus safe to assume that both stars are synchronized.

To summarize, the A star HD~74423 (TIC~355151781) shows an oblique pulsation aligned with the tidal bulge created by its binary companion. The single pulsation mode generates a frequency multiplet split by the orbital frequency, and the frequency and phase patterns demonstrate that the pulsation axis lies near to the line of apsides. HD~74423 is the first obliquely pulsating star known where the location of the pulsation axis is governed by the tidal distortion. Furthermore, and most interestingly, the pulsating star does so almost exclusively in only one hemisphere. It remains unclear whether this is the hemisphere facing its companion or the one pointing away from it.

The pulsation mode in HD~74423 is currently unique, but there must be a class of such stars that have their pulsation axes aligned with their tidal axes, and this discovery is an impetus to search for more. For instance, the pulsating primary of the eclipsing system U Gru (ref.$^{30}$) could be a related object. HD~74423 motivates more detailed studies of the interaction between stellar pulsations and tidal distortion in binary stars. We know from the heartbeat stars that tides do excite modes that are nearly resonant with orbital harmonics. Do tides in binary stars impact the excitation of p~modes and g~modes? While this question remains a puzzle, HD~74423 and other stars like it (when they are found) may be the key.

\subsection*{Acknowledgements}This paper includes data collected by the {\it TESS} mission. Funding for the {\it TESS} mission is provided by the NASA Explorer Program. Funding for the {\it TESS} Asteroseismic Science Operations Centre is provided by the Danish National Research Foundation (Grant agreement no.: DNRF106), ESA PRODEX (PEA 4000119301) and Stellar Astrophysics Centre (SAC) at Aarhus University. Some of the observations reported in this paper were obtained with the Southern African Large Telescope (SALT). Polish participation in SALT is funded by grant No. MNiSW DIR/WK/2016/07. DWK acknowledges financial support from the STFC via grant ST/M000877/1. MS is supported by an Australian Government Research Training Program (RTP) Scholarship. GH, SC, FKA and PS acknowledge financial support by the Polish NCN grant 2015/18/A/ST9/00578. DJ acknowledges support from the State Research Agency (AEI) of the Spanish Ministry of Science, Innovation and Universities (MCIU) and the European Regional Development Fund (FEDER) under grant AYA2017-83383-P. We thank the {\it TESS} team and staff and TASC/TASOC for their support of the present work and Allan R. Schmitt for making his light curve examining software {\tt LcTools} freely available. SC is grateful to Chris Engelbrecht for introducing him to the use of the observing equipment. GH thanks Ernst Paunzen for helpful discussions on the spectra of $\lambda$~Bootis stars.
\subsection*{Author contributions}GH provided the initial astrophysical interpretation for this object, coordinated the scientific analysis, analysed the photometric and spectroscopic data as well as oversaw and contributed to the paper writing. DWK carried out the frequency analysis and provided the interpretation in terms of the oblique pulsator model. SAR initiated the collaboration, oversaw and homogenized all aspects of the scientific analysis. HS modelled the pulsation amplitude and phase behaviour over the orbit. JF contributed the theoretical interpretation. DJ and PS analysed the ellipsoidal variability. ZG provided expertise in the modelling of pulsators in close binary systems. SC and FKA carried out auxiliary observations and computations. MS and SJM independently noticed the star's behaviour and provided their expertise. RG and TLJ originally pointed out the star to SAR and AV who work with the citizen scientists to vet their findings.
\subsection*{Competing Interests}The authors declare that they have no
competing financial interests.
\subsection*{Correspondence}Correspondence and requests for materials should be addressed to Gerald Handler.~(email: gerald@camk.edu.pl).

\section*{Methods}
\subsection*{Selection of interesting variable stars}
For the searches of unusual variables in {\it TESS} data the visual surveyors utilized the {\tt LcTools4} software ({https://sites.google.com/a/lctools.net/lctools/}, ref.$^{1}$).

\subsection*{{\it TESS} data, post-processing and analysis}
{\it TESS} 2-minute cadence data are delivered to the Mikulski Archive for Space Telescopes (MAST, {https://mast.stsci.edu/portal/Mashup/Clients/Mast/Portal.html}) in reduced form. The Pre-search Data Conditioning Simple Aperture Photometry (PDCSAP) flux data, converted to magnitudes, were used and analysed using a Discrete Fourier Transform (ref.$^{2}$) to produce amplitude spectra.

\subsection*{SALT data, post-processing and analysis}
A spectrum of HD 74423 was acquired with the Southern African Large Telescope (SALT) on the night of October 26/27, 2019 (HJD 2458783.59873). Reduced data by the SALT science pipeline (ref.$^{3}$, {http://pysalt.salt.ac.za/}) were used in this work. Only continuum normalization and removal of bad points was applied. A model atmosphere was computed with the program \texttt{SPECTRUM} (ref.$^{4}$, {http://www.appstate.edu/\~{}grayro/spectrum/spectrum.html}) and ATLAS9 model atmospheres (ref.$^{5}$) for purposes of illustration, and is compared with the observations in Supplementary Figure 1.

\subsection*{Data availability} TESS photometric data are publicly available at the Mikulski Archive for Space Telescopes (MAST, {https://mast.stsci.edu/portal/Mashup/Clients/Mast/Portal.html}). All relevant data are also available by request from the corresponding author.
\subsection*{Code availability} The codes for computing the Discrete Fourier Transform and to carry out the variability analyses are available on request from DWK. The \texttt{SPECTRUM} code used to compute synthetic spectra is publicly available from {http://www.appstate.edu/\~{}grayro/spectrum/spectrum.html}.

\newpage

\subsection*{Supplementary Figure 1}

\begin{figure}[ht!]
\centering
\includegraphics[trim=1.5cm 1.1cm 1cm 7.2cm,width=0.99\textwidth]{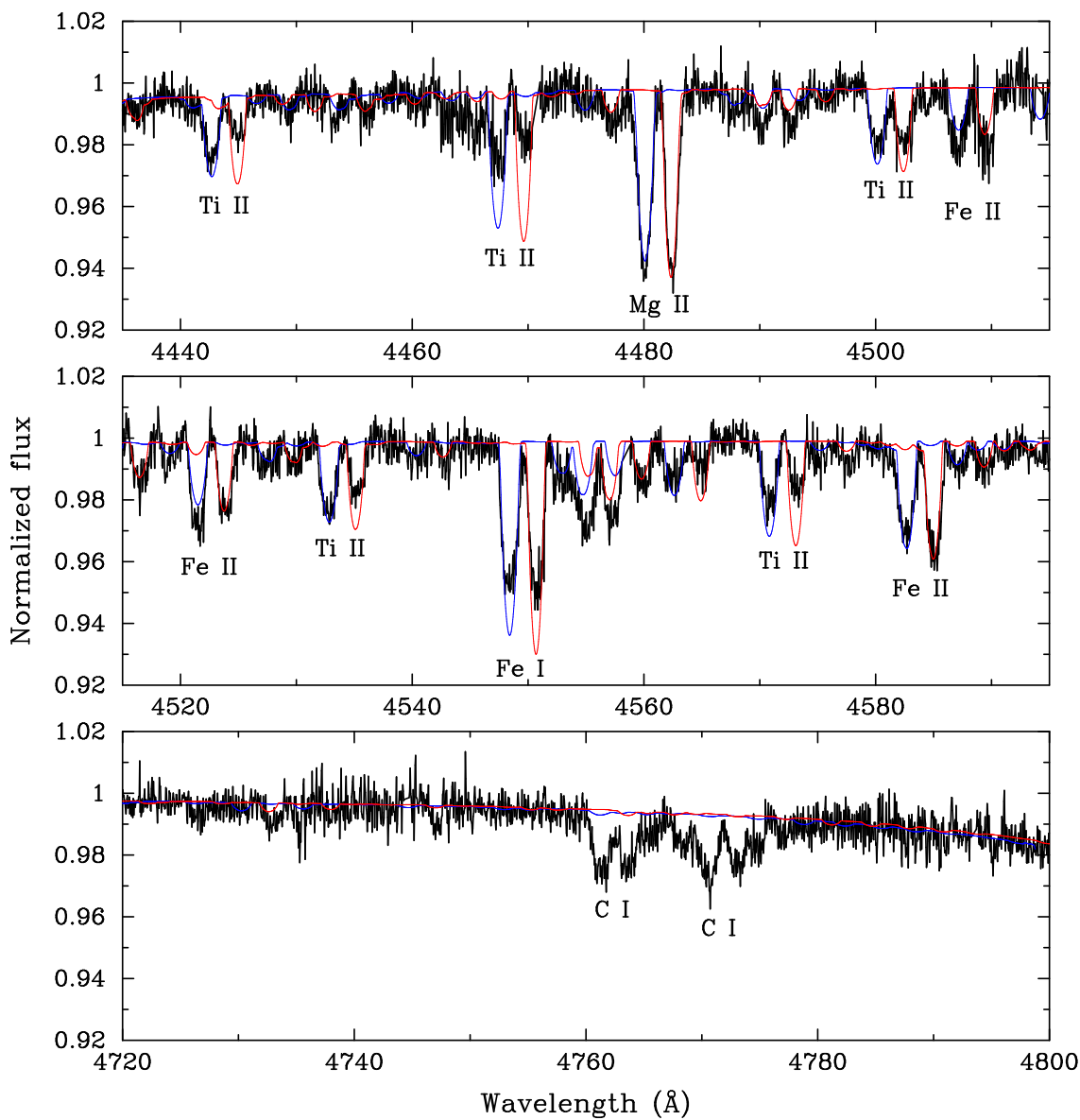}
\caption{Selected regions of a high-resolution spectrum of HD 74423 demonstrating it to be composed of two $\lambda$ Bootis stars. Top two panels: a region containing prominent metal lines of A-type stars. Lower panel: a region around a multiplet of carbon lines. The black graph is the observed spectrum, and the blue and red lines are two theoretical spectra with $T_{\rm eff}=8000$\,K, log $g=4.0$, $[M/H]=-2.0$ and $v \sin i=55$ and $50$\,\kms, respectively. The spectral lines are all double, with a separation of about $150$\,\kms. The strengths of most metal lines are reasonably well reproduced with such a low overall metallicity, with the exception of carbon that has a much larger abundance. Both components of HD 74423 thus share the $\lambda$ Bootis type spectral peculiarity.}
\label{fig:spec}
\end{figure}

\subsection*{Supplementary Figure 2}

\begin{figure}[ht!]
\centering
\includegraphics[width=0.95\textwidth]{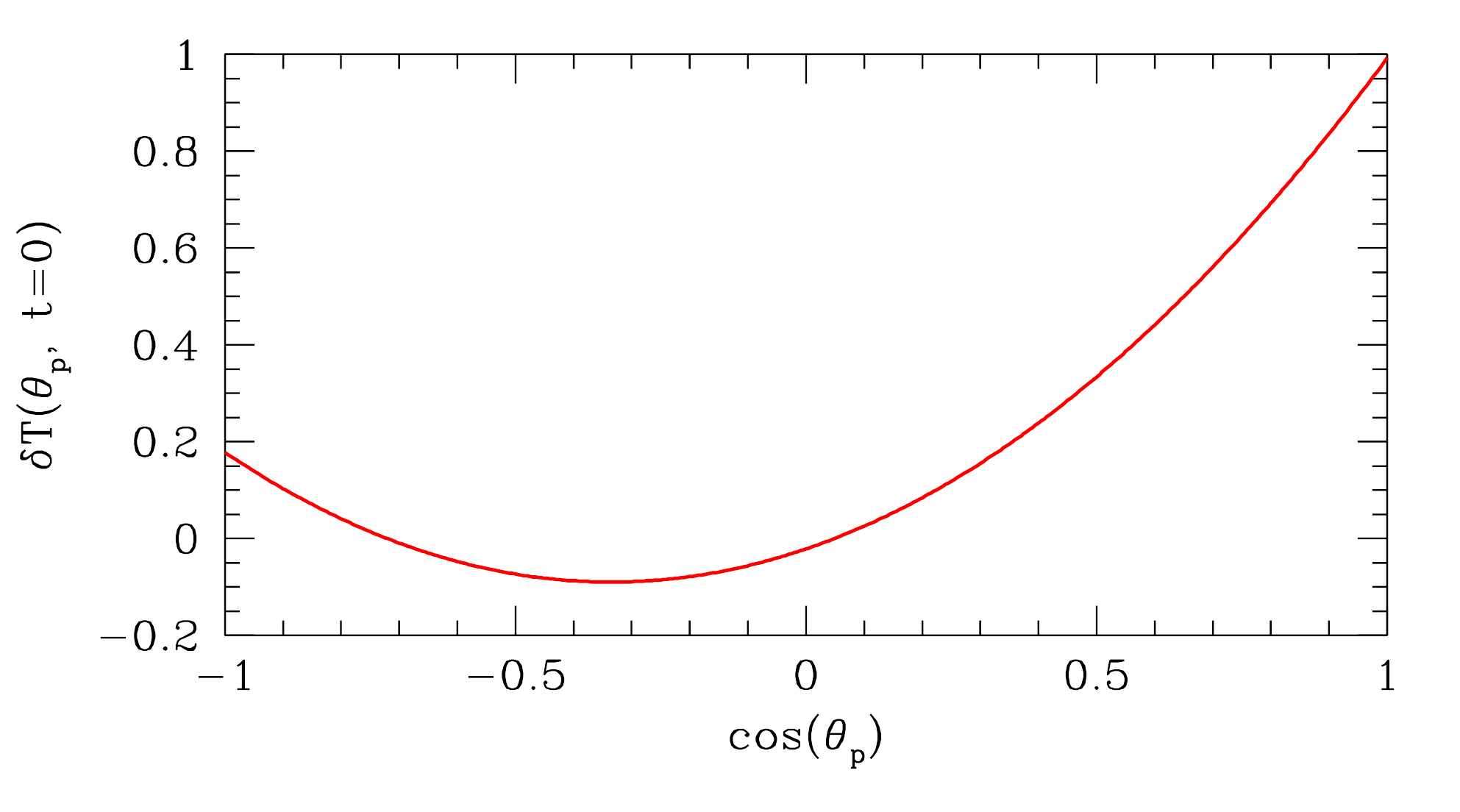}
\caption{Temperature perturbation at the initial pulsation phase as a function of $\cos\theta_{\mathrm p}$. This asymmetric distribution is required to reproduce the run of the pulsation amplitude in Fig.\,4 and indicates the pulsation amplitude to be trapped in one hemisphere.}
\label{fig:deltT}
\end{figure}

\end{document}